\documentclass[preprint]{aastex}
\usepackage{amsmath,amssymb,graphicx,longtable,lscape}

\newcommand{\tmop}[1]{\ensuremath{\operatorname{#1}}}

\newcommand{\tmtextit}[1]{{\itshape{#1}}}

\begin{document}

\title{Density of warm ionized gas near the Galactic Center: Low radio frequency observations}
\author{Subhashis Roy}
\affil{NCRA-TIFR, Pune-411007, India}
\email{roy@ncra.tifr.res.in}

\begin{abstract}
We have observed the Galactic Center (GC) region at 0.154 and 0.255 GHz with
the GMRT. A total of 62 compact likely extragalactic sources are detected.
Their scattering sizes go down linearly with increasing angular distance from
the GC up to about 1$^{\circ}$. The apparent scattering sizes of sources are
more than an order of magnitude down than predicted earlier by the NE2001 model
of Galactic electron distribution within 359.5$^{\circ} < l < 0.5 ^{\circ}$
and $-0.5^{\circ} <b <0.5^{\circ}$ (Hyperstrong scattering region) of the
Galaxy. 
High free-free optical depths ($\tau$) are observed towards most of the
extended nonthermal sources within 0.6$^{\circ}$ from the GC. Significant
variation of $\tau$ indicate the absorbing medium is patchy
at an angular scale of $\sim$10$'$ and n$_e$ is $\sim 10.$cm$^{-3}$ that
matches with the NE2001 model. This model predicts the extragalactic (EG)
sources to be resolved out from 1.4 GHz interferometric surveys.  However,
8 likely EG sources out of 10 expected in the region are present in 1.4
GHz catalog.  Ionized interfaces of dense molecular clouds to the ambient
medium are most likely responsible for strong scattering and low radio
frequency absorption.  However, dense GC clouds traced by CS $J=1-0$ emission
are found to have a narrow distribution of $\sim0.2^{\circ}$ across the
Galactic plane.  Angular distribution of most of the EG sources seen through
the so called Hyperstrong scattering region are random in $b$, and typically
$\sim$7 out of 10 sources will not be seen through to the dense molecular
clouds, and it explains why most of them are not scatter broadened at 1.4 GHz.

\end{abstract}

\keywords{Galaxy: center -- ISM: clouds -- radio continuum: ISM -- H II regions}

\section{Introduction:}

The central few hundred parsec region of the Galaxy is characterized by a
dense and turbulent interstellar medium, where the density and
velocity widths of spectral lines could almost be an order of magnitude higher
than the disk of the Galaxy. This environment prefers massive stars to form,
which provide copious amount of ultraviolet radiation causing ionization of molecular
clouds and thereby producing large column density of warm ionized medium
(WIM). This WIM causes high dispersion and scattering to the radiation
passing through it. Dispersion measure causes a frequency dependent delay,
while scattering due to electron density irregularities cause angular
broadening of sources viewed through it.

Dispersion is quantified by dispersion measure and can be measured
towards time variable sources like pulsars. It directly provides column
density of electrons towards that line of sight. At low radio frequencies,
free-free absorption by WIM also causes a frequency dependent absorption
to the radiation passing through it \citep{ANANTHARAMAIAH1991}, that provides
important information on the emission measure of this gas that can be used to
estimate the density of the WIM. Scattering due to WIM provides a
measure of turbulence in the medium and is related to the square of the
electron density fluctuations in the medium (see e.g., \citet{TAYLOR1993}).
Using pulsar dispersion measures, \citet{TAYLOR1993} made a model of electron
density distribution in our Galaxy that provides a valuable tool in providing the
distribution of the WIM in the Galaxy and a distance estimate of pulsars from
their dispersion measures. However, due to lack of pulsars discovered close to
the Galactic Center (GC), modeling electron density distribution near it
was not possible. Later, \citet{LAZIO1998} predicted the electron
density in the central $\sim 100$ pc to be about 10 cm$^{- 3}$. Their model is
largely based on (i) scatter broadening of Galactic sources (e.g., masers seen
towards the GC have angular sizes ranging from a few hundred milli arc-sec
\citep{VANLANGEVELDE1992} to about 1$''$ for Sgr A* at 1 GHz) and (ii) free-free
absorption towards the GC sources like Sgr A complex. They used scattering
sizes of a few known extragalactic sources to limit the angular size of the
scattering screen. This model has been incorporated in the improved model of
Galactic electron density distribution, NE2001 \citep{CORDES2004}. Since the
Hyperstrong scattering screen in this model is shown to be close to the GC, it
is inefficient in scattering the GC sources.  This model predicts scattering
size of extragalactic sources seen through it as $\sim 100''$ at 1 GHz, but
prediction of scattering diameter of EG sources based on scattering of GC
sources goes inversely proportional to the distance of the screen from the
GC.  Moreover, Galactic masers could be preferentially located in dense clouds,
the ionized surface of which could give rise to scattering, and thereby would introduce
error in scattering seen towards directions away from the clouds. Scattering size
of extragalactic sources are required to check the model prediction.

\citet{LAZIO1999} found the radio galaxy G359.87+0.18 seen through the
Hyperstrong scattering regime at 0.33 GHz with a scattering size of $\sim
20''$, that showed our lack of knowledge on filling factor of the Hyperstrong
scattering screen. Electron density is shown to drop sharply by two orders of
magnitude just outside the Hyperstrong scattering region. This sudden drop in
electron density is unrealistic and results from lack of constraints in NE2001.
Indeed, \citet{BOWER2001} have shown the scattering sizes of 3 extragalactic
sources close to but outside the Hyperstrong scattering screen to be
significantly larger than the \citet{TAYLOR1993} model. Moreover, In the last
one decade, some of the constraints used in the model has changed. For example,
emission from Sgr A* is shown not to undergo free-free absorption due
to Sgr A West below 950 MHz, but has been observed at much lower frequencies
\citep{NORD2004,ROY2004}.  Also, the GC region has been mapped at low radio
frequencies of 0.154 \citep{ROY2009} and 0.074 GHz \citep{BROGAN2003}.
Free-free absorption of sources seen through the GC region at these low
frequencies could better constrain the density of the WIM.

Scattering size is roughly proportional to the square of the observing
wavelength. Therefore, low radio frequency observations of the central
2$^{\circ}$ region of the GC could provide more detection of background
extragalactic sources, and measuring their angular sizes at more than one
wavelength would provide a measure of their true scattering size. We have
carried out Giant Metrewave radio telescope (GMRT) observations of this region
at 0.255 and 0.154 GHz. Preliminary results of these observations were published in
\citet{ROY2006,ROY2009}. Here, we present more detailed analysis from these
data sets and examine existing higher frequency observations in the region to
find the properties (i.e., electron density, angular extent and
filling factor) of the WIM in the central 2$^{\circ}$ region of the GC.

The remaining sections are arranged as follows - In sect. 2, observing and
data analysis procedures are described. In Sect. 3 and 4, we describe the
results obtained and discuss the implications respectively. Conclusions are
presented in Sect. 5.

\section{Observations and data analysis:}

The GC region was observed with the GMRT on 13th March 2003 at 0.225 GHz and on
20th August 2005 at 0.154 GHz with Pointing centers at RA (J2000)
=17h46m00$^{''}$ and DEC (J2000)=$-$28$^{\circ}$57$'$00$^{''}$ and at RA
(J2000) =17h45m40$^{''}$ and DEC (J2000)=$-$29$^{\circ}$00$'$28$^{''}$
respectively. The corresponding full width half maximum (FWHM) of the primary
beam at the above 2 frequencies are close to 126$'$ and 180$'$ respectively.
Sources located beyond the FWHM will be highly attenuated by the primary beam and
will be missed in our maps. The resolution of the whole array for a southern
source like the GC at the above two frequencies are $\sim 15^{''}$ and $\sim
25^{''}$ respectively.

3C48 was used as the absolute flux density calibrator, while 1830$-$36 and
1714$-$252 were used as secondary calibrators.
Observations at GMRT in the above two bands are often affected by radio
frequency interference (RFI), and to reduce it, a narrow bandwidth of 6 MHz
with relatively less RFI were chosen for both the observing frequencies. To
minimize any saturation of the antenna based electronics system in presence of
strong RFI, 14 db Solar attenuator at the pre-amplifier stage was used. In the
absence of System temperature measurement at GMRT, we have corrected for the
variation of Sky temperature from the calibrator to the target source following
\citet{ROY2004} keeping the automatic level control (ALC) off.  

After calibration and editing, a pseudo-continuum data base was made to
optimize for data size while ensuring bandwidth smearing during imaging smaller
than the synthesized beam up to half power point. The initial images were
improved by phase-only self-calibration (self-cal). To detect small diameter
sources in the field at 0.255 GHz, we first made a map of the full field using
polyhedral imaging in AIPS, and subtracted the Fourier transform of Clean
components of Sgr A complex and re-imaged the field with a short \tmtextit{uv}
cutoff to resolve out any extended emission of size larger than 7$'$. The
synthesized beam size of the resulted map is 17$^{''}$.3$ \times$12$^{''}$.5. The rms
noise achieved is $\sim$8 mJy.beam$^{- 1}$. This method was also used for
imaging the compact sources at 0.154 GHz. The short \tmtextit{uv} cutoff used
was 600 lambda, that resolves out structures with sizes more than 5$'$ (it also
avoids RFI seen in shorter baselines.  The synthesized beam size of the
resulted map is 33.4$'' \times$25.6$''$. The rms noise achieved is $\sim$10
mJy.beam$^{- 1}$, within a factor of two of the thermal noise.

To determine the sizes and flux densities of the compact sources, we deconvolved
the sources from the synthesized beam using the AIPS task JMFIT. It allowed us
to fit and subtract any background emission. We also accounted for bandwidth
smearing and primary beam attenuation in JMFIT to get a corrected source size
and flux density. The resultant catalog of compact sources is presented below.

\begin{figure}[htb]
{\includegraphics[width=0.7\textwidth,clip=true,angle=270,totalheight=10cm]{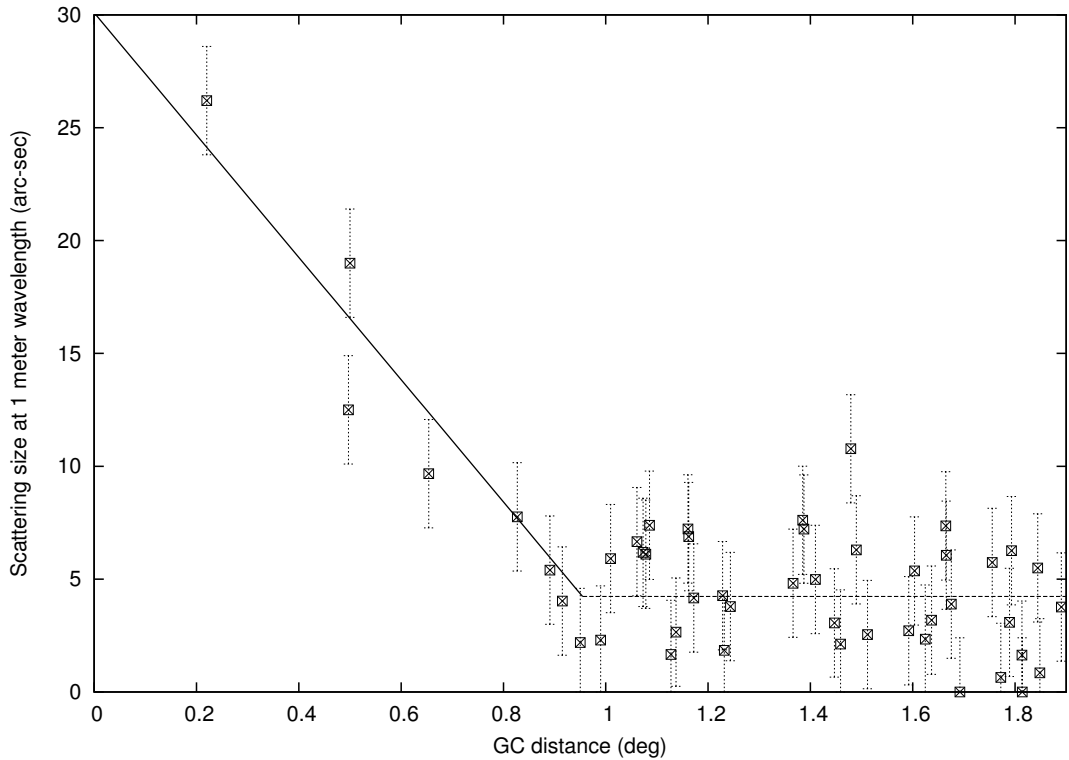}}
\caption{(a) Scattering size at a fiducial wavelength of 1 meter of likely
extragalactic (EG) sources comprising (i) seen within about 1$^{\circ}$ of the GC and
(ii) sources detected at 0.154 GHz between 1$^{\circ}$ to 1.9$^{\circ}$ from
the GC. The two straight lines (one continuous and the other dashed) show
multi-branch fits to the scattering data with (i) 0 $<$ GC distance $<$
0.95$^{\circ}$, and (ii)  0.95$^{\circ}$ $<$GC distance $<$1.9$^{\circ}$}
\end{figure}

\section{Results:}

In Table 1, we have tabulated the positions of the small diameter sources in
Galactic longitude ($l$), latitude ($b$) and source flux densities at 0.154
(S$_{154}$) and 0.255 GHz (S$_{255}$) from the GMRT and at 0.33 GHz from VLA
data \citep{NORD2004}. Error on these flux densities have two parts, the first
one arises from map rms noise, and the second one arises from absolute
calibration errors. Absolute calibration errors at the above two GMRT
frequencies are $\sim$5$-$10\%.

   \begin{longtable}{|c|c|c|c|c|c|c|c|c|c|c|l|}
\caption{Flux density and source sizes near the GC} \\
\hline
\small
   Galac.           & Galac.          & S$_{154}$       & S$_{255}$       & S$_{1400}$        & Spect. & S$_{exp}$ & $\frac {S_{exp}}{S}$ & $\theta_{154}$ & $\theta_{255}$ & $\theta_{1400}$ & Scat \\
long. ($\it l $)    & lat. ($\it b $) & (mJy)           & (mJy)           & (mJy)             & index  & (.154    & (.154                 &  $('')$        & $('')$ & $('')$ & size  \\
                    &                 &                 &                 &                   &        &  GHz)    &  GHz)                 &                &        &        &  $('')$   \\
\hline
\endfirsthead

\hline
Galac.           & Galac.          & S$_{154}$       & S$_{255}$       & S$_{1400}$        & Spect. & S$_{exp}$ & $\frac {S_{exp}}{S}$ & $\theta_{154}$ & $\theta_{255}$ & $\theta_{1400}$ & Scat \\
long. ($\it l $) & lat. ($\it b $) & (mJy)           & (mJy)           & (mJy)             & index  & (.154    & (.154                 &  $('')$ & $('')$                & $('')$ & size  \\
                 &                 &                 &                 &                   &        &  GHz)    &  GHz)                 &         &                       &        &  $('')$   \\
\hline
\endhead

357.864 & -0.994 & 1578 & --   & 882   & -0.8 &  1613  &  0.97  & 28.6 & --    & 5.5 & 5. \\ 
357.884 &  0.007 &  657 & --   & 292   & -0.9 &   576  &  1.14  & 28.8 & --    & 1.9 & 6. \\ 
358.002 & -0.635 & 1061 & --   & 654   & -0.4 &   884  &  1.19  & 24.7 & --    & 0.0 & 6. \\ 
358.153 & -1.677 & 1895 & --   & 1464  & -1.5 &  4543  &  0.41  & 26.1 & --    & 6.6 & 5. \\ 
358.156 &  0.030 &  405 & --   & 271   & -1.0 &   576  &  0.70  & 23.8 & --    & 1.6 & 5. \\ 
358.591 &  0.045 &  691 & 204  & 352   & -1.0 &   748  &  0.92  & 21.6 & 0.0   & 0.0 & 4. \\
358.605 &  1.441 &  421 & --   & 287   & -0.8 &   525  &  0.80  & 22.5 & --    & 0.0 & 5. \\ 
358.613 & -0.034 &  309 & --   & 232   & --   &   232  &  1.33  & 31.3 & --    & --  & -- \\ 
358.637 & -1.159 &  488 & --   & 252   & -1.2 &   623  &  0.78  & 13.4 & --    & 4.3 & 2. \\ 
358.786 &  1.268 &  413 & --   & 245   & -0.8 &   448  &  0.92  & 24.9 & --    & 3.6 & 5. \\ 
358.848 &  0.160 &  438 & 197  & 154   & -1.0 &   327  &  1.33  & 29.9 & 10.0  & 2.3 & 6. \\ 
358.872 &  0.276 &  250 & 231  & 102   & -1.1 &   234  &  1.06  & 31.3 & 14.2  & 0.0 & 8. \\
358.891 &  1.409 &  402 & --   & 144   & -0.6 &   226  &  1.77  & 27.2 & --    & 2.3 & 6. \\ 
358.898 &  1.641 &  778 & --   & 182   & -0.6 &   286  &  2.71  & 38.9 & --    & 0.0 & 9. \\ 
358.916 &  0.072 & 1573 & 2170 & 1623  & -1.1 &  3724  &  0.42  & 32.1 & 13.0  & 1.0 & 8. \\
358.932 & -1.196 &  598 & --   & 222   & -0.9 &   437  &  1.36  & 23.3 & --    & 5.3 & 4. \\ 
358.946 &  1.236 &  167 & --   & 99    & -0.9 &   195  &  0.85  & 10.1 & --    & 0.0 & 2. \\ 
358.982 &  0.580 & 178  & 275  & 164   & -0.4 &   221  &  0.80  & 18.0 & 9.00  & 0.0 & 5. \\
359.544 & -1.145 & 272  & 350  & 171   & -0.9 &   337  &  0.80  & 8.02 & 9.4   & 6.1 & 1. \\
359.546 &  0.988 & --   & 1010 & 249   & -0.5 &   --   &  --    & --   & 44.6  & 1.8 & 31. \\
359.604 &  0.306 & --   & 255  & 155   & -1.1 &   --   &  --    & --   & 34.8  & 8.2 & 19. \\ 
359.626 &  1.314 & 298  & --   & 200   & -0.8 &   365  &  0.81  & 20.9 & --    & 5.0 & 4. \\ 
359.707 & -0.585 & 289  & 171  & 129   & -0.8 &   236  &  1.22  & 47.5 & 17.2  & 4.7 & 10. \\
359.709 & -0.905 & 719  & 508  & 441   & -1.0 &   938  &  0.76  & 25.1 & 10    & 11. & 3. \\ 
359.776 &  1.988 & 754  & --   & 277   & --   &   277  &  2.72  & 28.0 & --    & --  & -- \\ 
359.844 & -1.842 & 1077 & --   & 764   & -0.8 &  1397  &  0.77  & 3.6 & --     & 2.3 & 0. \\ 
359.870 &  0.178 & --   & 442  & 508   & -0.7 &  --    &  --    & --   & 40.5  & 3.8  & 26. \\ 
359.911 & -1.812 & 452  & --   & 660   & -0.7 &  1119  &  0.40  & 0.00 & --    & 1.0 & 0. \\ 
359.986 &  1.385 & 346  & --   & 178   & -1.3 &   474  &  0.72  & 33.0 & --    & 1.5 & 7. \\ 
359.993 &  1.592 & 278  & --   & 214   & -0.7 &   363  &  0.76  & 11.7 & --    & 0.0 & 3. \\ 
0.003 & -0.891 & 1053  & 724 & 568 & -0.5 &   828  &  1.27 & 27.5 & 17.0 & 7.0 & 5. \\ 
0.121 &  0.018 & --    & 147 & 92  &  --  &   --   &   --  & --   & 17.8 & --  & -- \\
0.130 & -1.066 &  221  & 231 & 129 & -2.0 &   583  &  0.37 & 26.8 & 22.4 & 1.7 & 6. \\ 
0.271 &  1.197 &  693  & 388 & 259 & -1.1 &   594  &  1.16 & 18.5 & 10.0 & 0.0 & 5. \\
0.303 &  0.394 &  171  & 94  & 51  & -0.6 &    80  &  2.13 & 59.1 & 15.5 & 1.2 & 13. \\
0.313 &  1.646 &  661  & --  & 404 & -1.3 &  1078  &  0.61 & 16.9 &  --  & 4.8 & 3. \\ 
0.403 &  1.062 &  442  & 400 & 261 & -0.8 &   477  &  0.92 & 11.5 & 10.0 & 2.7 & 3. \\ 
0.409 &  0.979 &  195  & --  & 127 & --   &   127  &  1.53& 28.9  & --   & --  & -- \\ 
0.442 &  0.588 &  206  & 145 & 117 & -0.3 &   146  &  1.40 & 18.4 & 13.5  & 13.& 1. \\ 
0.447 &  0.592 &  106  & 110 & 95  & --   &    95  &  1.11& 08.4  & 10.2  & -- & 1. \\ 
0.560 & -0.817 &  146  & 111 & 120 & -0.9 &   236  &  0.61 & 17.9 & 6.0  & 5.2 & 2. \\ 
0.634 &  1.540 &  244  & --  & 89  & -0.7 &   150  &  1.61 & 26.3 & --    & 0.0 & 6. \\ 
0.654 &  1.055 &  344  & 244 & 181 & -0.9 &   357  &  0.96 & 16.4 & 12.0 & 4.2 & 4. \\ 
0.660 & -0.852 &  123  & 174 & 110 & -0.8 &   201  &  0.61 & 26.5 & 12.1 & 0.0 & 7. \\ 
0.665 & -0.035 &  196  & 156 & 195 &  1.4 &    67  &  2.89 & 24.9 & 12.8 & 6.4 & 5. \\  
0.720 &  0.407 &  244  & 122 & 87  & --   &    87  &  2.80& 33.8  & 11.0 & -- & 8. \\ 
0.722 &  1.303 &  185  & --  & 82  & --   &    82  &  2.25& 27.3  & --    & -- & -- \\ 
0.736 & -1.462 &  224  & --  & 169 & -0.8 &   309  &  0.72 & 13.8 & --    & 0.0 & 3. \\ 
0.799 & -1.794 &  398  & --  & 308 & --   &   308  &  1.29 & 32.3 & --    & 0.0 & 7. \\ 
0.846 &  1.174 &  707  & --  & 421 & -0.7 &   714  &  0.98 & 13.2 & --    & 1.5 & 3. \\ 
0.871 & -0.282 &  269  & 261 & 142 & -0.9 &   280  &  0.96 & 27.4 & 17.4 & 11. & 4. \\ 
0.899 & -1.401 &  217  & --  & 60  & -0.9 &   118  &  1.83 & 31.9 & --    & 4.6 & 6. \\ 
1.009 &  0.027 &  188  & 156 & 133 & -0.7 &   225  &  0.83 & 25.6 & 16.0 & 4.4 & 6. \\ 
1.060 &  0.382 &  160  & 181 & 125 & -1.1 &   286  &  0.55 & 7.22 & 4.0  & 0   & 2. \\ 
1.027 & -1.109 & 1342  & --  & 892 & -1.2 &  2207  &  0.60 & 11.0 & --    & 1.6 & 2. \\ 
1.047 &  1.574 &  633  & --  & 364 & -0.6 &   572  &  1.10 & 16.3 & --    & 2.1 & 3. \\ 
1.188 & -1.315 &  406  & --  & 327 & -0.7 &   554  &  0.73 & 2.7  & --    & 0.0 & 1. \\ 
1.408 & -0.382 &  246  & --  & 154 & -0.6 &   242  &  1.01 & 9.2  & --    & 3.4 & 1. \\ 
1.460 &  0.236 &  999  & --  & 237 & -1.2 &   586  &  1.70 & 46.8 & --    & 3.1 & 10. \\ 
1.479 & -0.822 &  197  & --  & 129 & -1.3 &   344  &  0.57 & 0.0  & --    & 3.2 & 0. \\ 
1.539 & -0.959 & 1012  & --  & 1099& -3.0 & 10585  &  0.09 & 7.0  & --    & 0.0 & 2. \\ 
1.826 &  1.071 &  678  & --  & 376 & -0.9 &   741  &  0.91 & 16.5 & --    & 1.0 & 4. \\ 
%
\hline
\multicolumn{12}{l}{Note: In the above table, `Galac.' indicates Galactic, S$_{154}$, S$_{255}$ and S$_{1400}$ -- flux densities }\\ 
\multicolumn{12}{l}{of sources at 0.154, 0.255 and 1.4 GHz respectively. `Spect.' -- spectral,} \\
\multicolumn{12}{l}{S$_{exp}$ -- expected flux density from extrapolation of spectral index between 0.33 }\\ 
\multicolumn{12}{l}{and 1.4 GHz and flux density at 0.33 GHz,} \\
\multicolumn{12}{l}{$\frac {S_{exp}}{S}$ -- ratio of expected to measured flux densities at 0.154 GHz,} \\
\multicolumn{12}{l}{$\theta_{154}$, $\theta_{255}$ and $\theta_{1400}$  -- angular sizes of sources at 0.154, 0.255 and 1.4 GHz respectively.}\\
\multicolumn{12}{l}{Scat size -- Scattering size at a fiducial wavelength of 1 meter.} \\
\multicolumn{12}{l}{Typical error on the size is $\sim$2.4$^{''}$.}
\end{longtable}

All the sources detected at 0.154 or 0.255 GHz were found to have counterparts
at 0.33 GHz survey \citep{NORD2004}. We compute spectral indices for all the
0.33 GHz sources based on detection of counterparts in the Galactic
plane survey of radio sources at 1.4 GHz \citep{WHITE2005} (hereafter, GPSR),
or in the survey of compact sources towards the GC \citep{LAZIO2008}
(hereafter, LC08).  Spectral indices ($\alpha$) for these sources have
been computed following the convention of flux density (S) $\propto
\nu^{\alpha}$  between 1.4 and 0.33 GHz. Observing time per field of GPSR was
much shorter than that of LC08 and they tried to maximize
$\tmop{uv}$-coverage by using multiple snapshot mode. Therefore, whenever
possible, we have used LC08 measurements rather than GPSR in this paper.
Deconvolved source sizes at 0.154, 0.255 and 1.4 GHz are quoted in Table 1.
Fractional errors on major axis size from deconvolution is in general
significantly lower than the minor axis. Since we are attempting to measure
scatter broadening of sources, rather than using the average size from the
above two measurements, we only quote the major axis size in Table-1.  In the
case of 0.33 GHz GC catalog of sources by \citet{NORD2004}, they quoted the
average source sizes rather than the major axis size and we have not used their
source sizes in estimating scattering sizes of the sources. Typical error on
the major axis sizes of the sources shown in Table-1 is about 2.4$^{''}$.

In \citet{ROY2009}, 26 sources were found from the 0.255 GHz GMRT map, and
the scattering diameter at a fiducial wavelength of 1 meter utilizing other
multi-frequency observations showed a decrease with angular distance from the
GC. To check that trend for all the 57 sources in Table 1 having observations
at 2 or more frequencies, we have fitted a $\lambda^{11/5}$ curve to their
measured angular sizes assuming a Kolmogorov wavenumber spectrum for the
electron density fluctuation \citep{TAYLOR1993} along with an additional
frequency independent term accounting for their intrinsic sizes ($a$)
[$\theta \left( \lambda \right) = a + b.  \lambda^{11/5}$, where $b$ is
scattering size at an wavelength of 1 meter]. 
For one of the sources G0.447+0.592, no 1.4 GHz detection exists,
and without it, the fit could not produce sensible result of scattering size. 
Another source, G0.442+0.588, the angular size and flux density as measured by
LC08 and in GPSR differs widely. Therefore, we have not used the
scattering sizes of the above two sources any further in the rest of this paper.
We have quoted the scattering sizes (derived value of $b$ from the fit) of the
sources for an wavelength of 1 meter in Table-1. 

\subsection{Scattering sizes and angular distances of sources from the GC}

Any dependence of scattering size of extragalactic sources with angular
separation from the GC would indicate presence of a scattering screen in the
region that scatters the emission from the sources. For all the source in the
list except three, the estimated spectral indices are consistent with their
extragalactic origin. The spectral indices of the three sources G0.130$-$1.066,
G0.665$-$0.035 and G1.539$-$0.959 are $-$2.0, 1.4 and $-$3.0 respectively,
which are atypical for extragalactic sources. Since scattering size depends
strongly on the distance to the source from the screen, and the above 3 sources
could be Galactic or variable sources, we have not used their scattering sizes
in the rest of the paper. 

From Table-1, we investigate for any correlation of scattering size with
angular distance from GC.  Spearman's rank correlation shows that there is a
strong anti-correlation of source angular distance from the GC with derived
scattering size for an wavelength of 1 meter for sources within about 1$^{\circ}$
from the GC. Spearman rank correlation coefficient is $-$0.98, indicating
anti-correlation with a confidence higher than 99.9\%.  As shown in Fig.
1, apparent scattering size appear to fall linearly from GC as a function of
angular distance. We have also plotted in Fig. 1, the scattering size of
sources detected at 0.154 GHz with angular distance between 1.0$^{\circ}$ to
1.9$^{\circ}$ from GC. As seen in the plot, a decrease in scattering size of
these sources as a function of angular
distance from GC is not evident. Therefore, we used multi-branch fitting (in
`gnuplot') and for minimum least square of residuals fitted $f(x)= A - B
\times x$ for sources within 0.95$^{\circ}$ of GC, and $f(x)= A - 0.95.B$ for sources
within 0.95$^{\circ}$ to 1.9$^{\circ}$ from GC. The fitted values are, $A=
30^{''} \pm 2^{''}.3$, and $B= 27^{''} \pm 2^{''}.5$. This shows the
scattering size of sources at an wavelength of 1 meter at the GC is about
30$^{''}$, and it falls down to about 4$^{''}$ about a degree away from it,
and stays to that value till about 1.9$^{\circ}$ from GC.

\subsection{Free-free absorption of sources}

Given the NE2001 model of the WIM in the central 1$^{\circ}$ of the GC and the
known form of free-free absorption by ionized gas at temperature $\sim$6000K,
we expect significant absorption of sources at 0.154 GHz which are located
beyond the GC. Given the spectral indices of sources between 1.4 and
0.33 GHz (assuming no absorption) in Table 1, we quoted the 
expected source flux densities (S$_{exp}$) at 0.154 GHz using the relation:
S$_{exp}$($\nu_1) = $ [S$(\nu_2) \times(\nu_1/\nu_2)^{\alpha}$] \\
where $\nu_1$ and $\nu_2$ refers to 0.154 and 0.330 GHz respectively.

If the ratio of S$_{exp}$ (0.154 GHz) to measured flux densities S (0.154 GHz)
in Table-1 is less than unity, and source variability is neglected, then it
provides a value of $e^{-\tau}$ at 0.154 GHz caused by free-free absorption.
We find the median value of S$_{exp}$/S at 0.154 GHz in Table-1 to be $\sim$0.9,
suggesting absorption. However, $\sim$30\% of the sources show $e^{-\tau}$ $>$1,
indicating either curvature in the spectral indices derived from higher
frequencies or variability of source flux densities. Therefore, we do not draw
significant conclusions from the above ratio of flux densities.

Multi epoch flux density measurements of extended sources in the GC region are
free from source variability.  Therefore, to constrain the properties of the GC
WIM through absorption, we have measured free-free optical depth towards
non-thermal extended sources assuming no significant synchrotron self
absorption of the sources within about a degree from the GC and is tabulated
in Table-2. These sources are either indicated
\citep{ROY2003,LANG2010,MEREGHETTI1998,KASSIM1996} or believed to be close to
the GC distance along our line of sight. The flux densities at 0.154 GHz (S$_{154}$)
were measured from our map that was made using \tmtextit{uv}-spacings such that
it is sensitive to sources with angular sizes of up to 45$'$.  The 0.33 GHz
flux densities (S$_{330}$) were measured from the map by \citet{LAROSA2000}.
Flux densities from both the maps were measured after subtracting a background
near the respective sources in each of the maps. The free-free optical depth
($\tau$) at 0.154 GHz was then computed by using their higher frequency
spectral indices ($\alpha$) following the relation: \\ S($\nu_1) =e^{-\tau}
\times $ [S$(\nu_2) \times(\nu_1/\nu_2)^{\alpha}$] \\ where $\nu_1$ and $\nu_2$
refers to 0.154 and 0.330 GHz respectively.

\begin{table}[!htb]
   \caption{Free-free optical depth towards the GC extended sources}
  \begin{tabular}{|c|c|c|c|c|}
\hline
Source name               & S$_{154}$ & S$_{330}$ & Spectral index & $\tau$ (0.154 GHz) \\
and part                  & (Jy)            & (Jy) &  ($\alpha$)  &                  \\
\hline
Sgr A East $^a$           &  20$\pm$4 & 90.0$\pm$10  & $-$0.5$\pm0.2$ & 2.3$\pm$0.3  \\
7$'$ halo $^a$            &  55$\pm$20 & 210.0$\pm$50 & $-$0.27 $\pm0.2$ & 1.8$\pm$0.5 \\
Radio-arc (east side) $^b$&  20$\pm$2 & 19.0$\pm$3   &   +0.3 $^c$   &$-$0.3$\pm$0.2  \\
Sgr-C (west side) $^d$    & 2.9$\pm$0.5 & 5.4$\pm$1  & $-$0.5 $^e$   &  1.$\pm$0.2 \\
SNR G00.3+0.0             &  23$\pm$3 & 35.0$\pm$3   & $-$0.56 $^f$  &  0.8$\pm$0.2 \\
NTF G359.79+0.17$~^i$     &  0.5$\pm$0.05 & 1.4$\pm$0.1 & $-$0.6 $^c$& 1.5$\pm$0.2 \\
NTF G359.54+0.18$~^h$     &  $<$0.2       & 1.4$\pm$0.1 & $-$0.6 $^c$& $>$1.5      \\
NTF G0.08+0.15$~^g$       &  0.73$\pm$0.1 & 2.$\pm$0.2 & $-$0.6 $^c$ & 1.5$\pm$0.2  \\
G0.9+0.1                  & 13$\pm$1.5 & 11.5$\pm$1.5 & $-$0.77 $^e$ &  0.4$\pm$0.2 \\
SNR Sgr-D                 & 26$\pm$2   & 21.2$\pm$2   & $-$0.4 $^e$  &  0.1$\pm$0.2  \\
NTF Pelican               & 1.0$\pm$0.2 & 0.67$\pm$0.1 & $-$0.6 $^j$ & 0.0$\pm$0.2  \\
\hline
\multicolumn{5}{l}{In the above table, $^a$ -- from \citet{ROY2009}, $^b$
denotes part of Radio-arc seen east of} \\ 
\multicolumn{5}{l}{G0.17$-$0.07 at 0.154 GHz, $^c$ -- \citep{ANANTHARAMAIAH1991}, 
$^d$ -- part of Sgr-C seen west} \\
\multicolumn{5}{l}{of G359.45$-$0.05 at 0.154 GHz, $^e$ -- \citep{LAROSA2000},
$^f$ -- \citep{KASSIM1996}}\\ 
\multicolumn{5}{l}{$^g$ -- seen at 0.154 GHz between G0.07+0.13 to G0.09+0.22} \\
\multicolumn{5}{l}{$^h$ -- not detected at 0.154 GHz and 3$\sigma$ upper limit
as its flux density}\\
\multicolumn{5}{l}{$^i$ -- seen at 0.154 GHz near its brightest part G359.796+0.16 and $^j$ -- \citep{LANG1999a}.}
  \end{tabular}
\end{table}

As can be seen from Table-2, part of Radio-arc seen $\sim$15$'$ away from the
GC show negative $\tau$. This is likely to be due to flattening of the 
intrinsic spectral index at meter wavelengths compared to what is determined at
higher frequencies. Despite that, no significant absorption is seen towards the
east side of the Radio-arc. Part of Radio-arc has also been seen in the VLA 74
MHz map of this region \citep{BROGAN2003}.  We note that the 0.154 GHz flux
density of the SNR G0.3+0.0 is found to be significantly lower than the works
referenced in \citet{KASSIM1996} for its flux density near 0.12 GHz. The 0.154
GHz map resolves out emission with size scale bigger than about 45$'$, and
hence our flux density is a lower limit. However, being much devoid of extended
Galactic plane contribution, our flux density could be more accurate than the
low frequency observations quoted there, which were sensitive to larger scale
structures.

\section{Discussions}
\subsection{Are the small diameter sources extragalactic ?}

We have detected 62 sources from the 0.154 GHz GMRT map. Intrinsic angular
size as evident from their 1.4 GHz deconvolved size (Table 1) for all except
two sources are well below 10$''$. For the remaining 2 sources, their angular
sizes are also $\sim 10''$. Intrinsic angular size of 10$''$ and total flux
density greater than 100 mJy at 0.154 GHz indicate their intrinsic brightness
temperature to be $>$50,000 K. This high a brightness temperature shows that
emission from these sources are non-thermal, and no Galactic HII region is
expected in the list. To categorize them to be behind the GC, HI
absorption by known GC anomalous velocity features \citep{COHEN1976} need to
be detected. However, given their flux density at 1.4 GHz, detecting HI absorption
would require a lot of observing time, and would be difficult to get with the
presently operational interferometers.  However, we can categorize them as
likely extragalactic objects if their properties do not match with known
Galactic nonthermal sources, but is consistent with extragalactic ones.  As
discussed in \citet{ROY2005}, there are five types of Galactic nonthermal
sources known, namely, (i) SNRs, (ii) radio stars, (iii) Galactic microquasars,
(iv) radio pulsars and (v) radio transients.  We follow the arguments in
\citet{ROY2005} and check if these sources could fall in to any of the five
type of sources.  Considering the source sizes and range of flux densities at
the given frequencies, we find that none of the sources are likely to be SNRs
and radio stars.

Galactic microquasars are generally detected from their X-ray emission.  We
have searched for any possible counterparts in X-ray surveys of the region as
available from NASA/IPAC extragalactic database (NED, available on-line through
http://ned.ipac.caltech.edu). For 5 of the sources in Table 1, we found X-ray
counterparts within 0.5$^{'}$ of the radio position. These are G358.891+1.409,
G359.911$-$1.812, G0.121+0.018, G0.722+1.303 and G0.665$-$0.035.
G358.891+1.409 is the radio counterpart of the X-ray source GRS 1734$-$292 and is
categorized as a Seyfert 1 galaxy behind the GC \citep{MARTI1998}.
G359.911$-$1.812 lies within the FWHM of a XMM detected X-ray source.  This
source cannot be ruled out as being Galactic. However, its scattering diameter
is found to be about 0$^{''}$, and being located $\sim 2^{\circ}$ away from the
GC, do not significantly bias the overall results presented here. G0.121+0.018
positionally matches with a Chandra detected X-ray source \citep{WANG2006}.
Given a high density of compact X-ray sources close to this region ($\sim$1 per
10$^{''}$), this could be a chance coincidence. However, scattering diameter
for this source could not be determined, and therefore its nature does not
affect any scattering results presented in this paper. The same conclusion is
drawn for G0.722+1.303, that has a possible X-ray counterpart, but its
scattering size could not be determined. For G0.665$-$0.035, the most accurate
coordinate is given in \citet{LAZIO2008} as G0.666$-$0.032, and the nearest
Chandra X-ray source is 20$^{''}$ away, that is much larger than their
positional uncertainty. We conclude that except one (G359.911$-$1.812), rest of
the sources with measured scattering diameters are highly unlikely to be
Galactic microquasars. 

Radio pulsars typically show a very steep spectral index with an average value of
$- 1.8 \pm 0.2$ \citep{MARON2000}.  In Table-1, there are only 4 objects whose
spectral indices determined from 1.4 and 0.33 GHz observations are steeper than
$-$1.2.  Therefore, we conclude that not more than a few objects in the above
list of sources could be pulsars.  

Given that all the sources in table-1 are detected in multiple wavebands, and the low rate
of finding transients after many hours of observations, these are also unlikely
to be of transient origin. While the possibility of some of these to be members
of new types of Galactic sources cannot be ruled out, but \citet{BECKER2010}
finds Galactic variable sources to have either a flat or inverted spectrum.
Except one source, there is no other source whose spectral index is flatter
than $-$0.3, and hence they are unlikely to be variable Galactic sources. 

We have estimated the number of extragalactic sources expected in the
0.154 GHz map from the known extragalactic source count at this frequency
\citep{INTEMA2011}. The lowest flux density in Table-1 at 0.154 GHz is 106 mJy.
The expected number of extragalactic sources above 100 mJy in the GMRT map with
correction for the primary beam is estimated to be about 70. Scattering
broadens the source size, thereby reducing the peak flux density, but it
preserves the total flux density. 
The map used to measure the scattering size at 0.154 GHz was sensitive to
structures of sizes $\sim 5'$. Therefore, sources which are scattered broadend beyond
this would not be detected. Moreover, if a point source is scatter broadened to
$\theta^{''}$ and given a resolution of $X^{''}$, its peak flux density will go
down by a factor of $\frac{(X^2+\theta^2)}{X^2}$. As $\theta$ increases more
than that of $X$, the peak flux density starts falling drastically. Unless the
source peak flux density of the source is $\sim 5 \times map~rms $, it may not be
detected. This would have greatly reduced detection of sources that are scatter
broadened to more than our resolution ($\sim 30^{''}$). However, as we have also
made maps at lower resolutions, brighter sources with scatter broadened sizes
of up to $\sim 1'$ would have been detected. Sources with scattering
sizes of more than 1$'$ could have been missed at 0.154 GHz. However, such
sources will have a scattering diameter a factor of 3 smaller at 0.255 GHz.
Therefore, sources with scattering sizes of up to $\sim 2-3'$ at 0.154 GHz and
seen within a degree from the GC (FWHM of the primary beam of GMRT at 0.255
GHz) are likely to be present in Table 1. From Fig. 1, estimated scattering
size nearly equals the synthesized beam size at 0.154 GHz at a distance of
about 0.7$^{\circ}$ away from the GC.  The expected mean scattering diameter in
the region is $\sim 60^{''}$ at 0.154 GHz. Consequently, source peak flux
density would be lower by a factor of 5. From the known EG source count and a
5$\sigma$ (rms $\sim$ 15 mJy.beam$^{-1}$ close to the GC) detectability of 0.75
Jy, in the absence of scattering we expect to see about 18 sources in the
region (radius 0.7$^{\circ}$). However, due to a factor of 5 reduction in peak
flux density (average), the expected source count goes down to about 5 this
region. We note that only 3 sources are detected in this region at 0.154 GHz.
However, given the statistical fluctuation of sources seen through each field
of view, the number of sources detected in the whole field of view at 0.154 GHz
are consistent with extragalactic source counts and we believe barring a few,
all the above sources are extragalactic.

\subsection{Implications of free-free absorption towards the GC sources}

Optical depth is related to the electron density ($n_e$) and temperature
($T_e$) as $\tau = \int 0.2 \times n_e^2 \times \nu^{- 2.1} \times T_e^{-
1.35} \tmop{dl}$, and emission measure is defined as $\int n_e^2 \tmop{dl}$.
In literature, temperature of WIM is typically taken to be 5000--10000
K. In this paper, we use 6000 K to be the temperature of the GC WIM.
Hence, emission measure of ionized gas responsible for a $\tau$ of
unity at 0.154 GHz is about 10$^4$ pc.cm$^{- 6}$ and it would decrease by a
factor of two in case the temperature of the WIM is 10,000 K. From
\citet{ROY2009}, we find the emission measure towards Sgr-A complex is $\sim
10^4$ pc.cm$^{- 6}$. From Table-2 (Sect. 3.2), we find that free-free
absorption of extended sources in the central 2$^{\circ}$ region of the GC is
quite common at 0.154 GHz. Higher values ($>$0.5) of $\tau$ were only obtained
for sources within 0.6$^{\circ}$ from the GC, where 7 out of 8 extended sources
show large absorption (Table-2). This suggests that the probability of
encountering ionized clouds whose emission measure is $\sim 10^4$~cm$^{-6}$.pc
along our line of sights within $\sim 0.6^{\circ}$ from the GC is high
($\sim$80\%).  $\tau$ values towards the NTF Pelican and the SNR Sgr-D seen
$\sim$1$^{\circ}$ away from the GC were much smaller.  While, it was moderate
towards the SNR G0.9+0.1 seen at similar angular distance from the GC,
suggesting a decrease in the emission measure of ionized clouds seen a degree
away from the GC. This indicates the free-free absorbing clouds to be mostly
located within $\sim$100 pc (corresponding to $\sim 0.6^{\circ}$ at the GC
distance) from the GC.

We find that $\tau$ vary widely towards different extended sources particularly
for the ones within the above 0.6$^{\circ}$ region from the GC. For example,
$\tau$ of the free-free gas towards the non-thermal filaments (NTFs)
G359.79+0.17 and G0.08+0.15 is significantly higher than the SNR G0.3+0.0.
$\tau$ is close to zero for eastern part of the Radio-arc. On close examination
of ratio map made by division of 0.154 GHz map by the 0.33 GHz map, we find
that $\tau$ varies smoothly by $\sim$0.5 across the length of the visible arc
extending more than 8$'$ ($\sim$20 pc). The same smooth variation of $\tau$ is
seen across the ratio map of size $\sim 5'$ centered on Sgr A East. These
suggest that the WIM responsible for absorption is not ubiquitous and it is not
concentrated in many small ($\lesssim 1'$) clouds.
From the above, we conclude that angular scale of absorption is $\sim$10$'$.
This corresponds to a length scale of $\sim$25 pc at the distance of the GC,
and indicates that absorption within 0.6$^{\circ}$ from the GC is
caused by ionized clouds whose emission measure is $\sim 10^4$~cm$^{-6}$~pc.
Their length scale implies electron density in the clouds to be $\sim$10 cm$^{-3}$. 

\subsection{Scattering medium}

Variation in plasma density scatters incoming rays from the source towards
slightly different directions after passing through the scattering medium. If
$\theta_s$ is the deviation of the direction of the incoming rays after
passing through the media, then for an extragalactic source, it is shown to be
empirically related to the scattering measure (SM) by the following relation:

$\tmop{SM} = \left( \frac{\theta_s}{128 \tmop{mas}} \right)^{\frac{5}{3}} .
\nu^{\frac{11}{3}}$, and SM is defined as $\tmop{SM} \equiv \int C^2_n
\tmop{dl}$. $C^2_n$ is the spectral coefficient for a truncated power law
spectrum of spatial fluctuations in the ISM electron density \citep{TAYLOR1993}. 
It is also shown that $d \left( \tmop{SM} \right) = C_u F n_e^2
\tmop{dl}$, where $C_u$ is a constant, and $F$ is a fluctuation parameter
relating SM to the square of the local volume averaged electron density
(n$_e$). The ratio of apparent scattering diameter of an EG source to a GC
source is given by $\frac {\theta_{s(EG)}}{\theta_{s(GC)}} = \frac{D}{d} $
\citep{LAZIO1998}, where the GC is located at a distance D from the observer,
and the screen is at a distance $d$ from the GC sources.
From our results (Fig. 1), we find the apparent scattering size of sources for
a fiducial observing wavelength of 1 meter decreases linearly within about
1$^{\circ}$ from the GC. If the electron density fluctuation parameter does not
change significantly in the region, it would indicate $\int{n_e^2.dl}$ to also
fall down in similar fashion. Since the peak of scattering is coincident with
the GC, it is unlikely to be a chance coincidence of a ionized cloud located
somewhere along the line of sight. Rather, it likely encompasses the GC region.
This indicates that indeed there is a GC component of scattering seen within
$\sim$1$^{\circ}$ from the GC. At the distance of the GC, its linear size is
$\sim$100 pc. However, the scattering size produced by this gas at 1
GHz is only $\sim$3$''$, more than an order of magnitude smaller than what is
expected from the Hyperstrong scattering screen. This could happen if the
electron density in this extended component of the WIM is almost an order of
magnitude smaller than predicted from the above model. It would produce an
emission measure of $\sim$500 pc.cm$^{- 6}$. Corresponding optical depth
produced by this screen will be less than unity at 0.074 GHz. If such an
extended ionized medium indeed exists, it would cause perceptible free-free
absorption at $\sim$0.035 GHz or less. Hence, ionized gas responsible for absorption
at 0.154 GHz is different from the extended ionized gas responsible for
scattering of the sources presented here. 

Two extragalactic sources, G359.870+0.178 and G359.604+0.306 seen through the
Hyperstrong scattering region,
but shows no large scattering (Table-1) as predicted by the Hyperstrong
scattering model. We have observed 7 objects seen through 0.5$^{\circ}$ from
the GC region. Other than the 2 compact sources, it includes eastern part of
Radio-arc that shows lack of absorption (Sect. 3.2), thereby indicating free-free
electron density at least by a factor of three lower than predicted by the
NE2001 model in that direction. Since NE2001 is the same as the Hyperstrong
scattering model in this region, it shows that out of 10 objects, this medium
has not been observed along three of them. We note that absorption towards 7
other objects only indicates presence of dense ionized cloud as required by the
NE2001 model. 
We conclude that probability of encountering Hyperstrong scattering medium
along our line of sights within $\sim 0.6^{\circ}$ from the GC is $\sim$70\%.
Due to small number of sources involved, the above probability only provides an
order of magnitude estimate of it.

\subsubsection{Hyperstrong scattering and missing sources}

The Hyperstrong scattering medium is expected to be present in the $ 359.5^{\circ} < l <
0.5^{\circ}$ and $- 0.5^{\circ} < b < 0.5^{\circ}$ region of the Galaxy and the
predicted scattering size of EG sources was $\sim$100$''$ at 1
GHz \citep{LAZIO1998}. This implied a scattering size of $\sim$4$^{''}$
for EG sources at 4.8 GHz, and is comparable to the resolution of the Galactic
plane survey of radio sources at 4.8 GHz \citep{WHITE2005} (hereafter, GPSR5).
Therefore, most of the background sources at 4.8 GHz will not be resolved out,
and is cataloged in the GPSR5. Assuming scatter broadening to be $\propto
\lambda^{11/5}$, this implies a scattering size of $\sim$50$''$ at 1.4 GHz.
Hence, all the EG sources seen through the region will be resolved out in
recent interferometric surveys of this region at 1.4 GHz (e.g., GPSR and
LC08). However, if the Hyperstrong scattering model is incorrect,
the ratio of scattering diameter between EG and GC sources could be much lower.
Since the resolution of the GPSR at 1.4 GHz is $\sim$5$''$, they would remain
in catalogs and allow us to differentiate between the above 2 cases. 

GPSR had a comparatively poor \tmtextit{uv}-coverage and due to a comparatively
large primary beam size of the VLA at 1.4 GHz, presence of intense emission
from Sgr-A East in the primary beam or through its first sidelobe causes
significant artifacts within $\sim 30'$ from the GC. Due to a much better
\tmtextit{uv}-coverage in maps made by LC08 at 1.4 GHz, we used their catalogs
to identify likely extragalactic sources at this frequency based on their steep
spectral index ($- 1.5 < \alpha < - 0.5$). In LC08, fifty eight sources are
seen through the Hyperstrong scattering medium, and the median flux density of
these sources is 43 mJy. We identified sources common to both GPSR5 and LC08 in
the region.  Twenty three sources/`source components' were identified, and at
1.4 GHz, their median flux density is 50 mJy.  To remove Galactic HII regions
based on their flat or inverted spectrum, we have determined the spectrum for
all these sources, and eliminated the sources suspected to be HII regions from
the list.  We have also visually checked the source positional coincidence with
other extended sources in the 0.33 GHz GC map \citep{LAROSA2000}. Whenever we
find any of the compact sources coincident on known extended sources, we have
removed them from the list as these could be part of that extended object.
After removal of these sources, nine sources from LC08 at 1.4 GHz positionally
matched with the GPSR5 or 0.33 GHz catalog of sources \citep{NORD2004}. These
are, G359.604+0.304, G359.718$-$0.036, G359.872+0.178, G359.775$-$0.449,
G0.209$-$0.002, G0.305+0.394, G0.352$-$0.068, G0.391+0.231 and G0.432+0.261.
These sources have spectrum corresponding to what is
expected for EG sources ($-1.5 < \alpha< -0.3$). 

To categorize them as EG objects, we need to check that these objects do not
belong to known classes of Galactic nonthermal sources.  Following the
discussion in Sect. 4.1, we consider their sizes, observing frequencies and
range of flux densities, and conclude that these sources are unlikely to be
SNRs, radio pulsars and radio stars. Galactic microquasars are generally
detected from their X-ray emission. We searched existing astronomical databases
using NED. An X-ray source was found to be
positionally coincident with G0.391+0.231. Therefore, it cannot be ruled out as
being a Galactic microquasar.  We note that all the above sources are detected
in multiple wavebands, and given the slow rate of finding transients after many
hours of observations, these are unlikely to be of transient origin. Therefore,
we believe except one possible exception (G0.391+0.231), all the other sources
in the above list of nine sources are extragalactic. 

We realized that using the lowest flux density in a sample of sources and using
that to estimate the EG source count in that frequency could be highly
prone to errors close to the GC, where (i) rms noise changes significantly
from one field to the other and (ii) the possibility of detecting fake sources
that could be caused by the diffraction pattern of a strong source seen
somewhere else near the field. Median flux density of a sample of sources is
less affected by these artefacts. Therefore, we used the median flux
density of these sources (50 mJy) at 1.4 GHz and estimated the number of
sources expected above that flux density. The expected number of sources seen
through 1 square degree is found to be about 5. Since the above number of
sources is estimated based on median, there would be an equal number of objects
below that flux density. Therefore, the total number of EG sources expected 
at 1.4 GHz in the catalog of the above region would be doubled, and is ten.
Since eight sources have been identified as possibly EG, we are
unlikely seeing the effect of Hyperstrong scattering for most of the sources.
This indicates that probability of encountering Hyperstrong medium along our
line of sights within $\sim 0.5^{\circ}$ from the GC is only $\sim$20\%.

\subsection{A model for the electron density distribution near the GC}

In Sect. 4.2, we found that significant absorption towards known extended
sources seen through $\lesssim 0.6^{\circ}$ from the GC region is quite common
and concluded that electron density in the clouds to be $\sim$10 cm$^{- 3}$.
Such an electron density was predicted by NE2001, and based on
conclusions drawn in Sect. 4.3, we expect to see Hyperstrong scattering of
$\sim$70\% EG sources in the region. However, in Sect. 4.3.1, we find that only
$\sim$20\% of the sources appear to have got resolved out at 1.4 GHz. 
\citet{LAZIO1998} discussed physical conditions in the scattering region in
detail, and the interface of dense molecular clouds to the ambient hot ionized
medium was preferred to explain the hyperstrong scattering in the GC region.
More recently, \citep{RODRIGUEZ2005} found the dense molecular clouds are
ionized by distant sources in the GC region. This indicates the surfaces of
clouds to be substantially ionized and will cause large scattering due to high
electron density in the region as compared to typical HII regions in Galactic disk.
Due to clumpiness of the clouds, the depth of ionization could be of the $\sim$pc
\citep{HOWE1991}. Consequently, the line of sights which passes close to any GC
molecular clouds would only be affected by strong scattering.

The GC molecular clouds are characterized by their high density and therefore
could be identified by CS $J= 1-0$ emission.  This emission was mapped by
\citet{TSUBOI1999}, and we have inspected the maps.  Their velocity integrated
map (Fig. 2(a)in \citet{TSUBOI1999}) shows that the GC molecular clouds have a
narrow scale height and near the GC, these clouds are typically seen within
$-0.1^{\circ} <b <0.2^{\circ}$. We note that most of the extended sources
within 0.6$^{\circ}$ of the GC in Table-2 were close to the Galactic plane.
However, the extragalactic sources mentioned in Sect. 4.3.1 are likely to be
randomly distributed in the region inspected within $-0.5^{\circ} <b
<0.5^{\circ}$. Therefore, the dense GC clouds will cause significant scattering
and absorption for sources only close ($\sim0.1-0.2^{\circ}$) to the Galactic
plane. This would then explain why most extragalactic sources are not found to
be missing from the 1.4 GHz catalogs made from interferometric observations.

We suggest a model where the Hyperstrong scattering medium is asymmetrically
distributed in $l$ and $b$. Its distribution along $l$ would be much more than
in $b$. Referring to the figure in \citet{TSUBOI1999}, the medium could be
extended up to $\sim \pm 1^{\circ}$ in $l$, but $\sim \pm 0.1^{\circ}-
0.2^{\circ}$ in $b$. Sgr-A* and the masers (Sect. 1) are likely seen through
this medium that causes significant scatter broadening to them. Following Sect.
4.3, there is a need to include an additional uniform moderate electron density
($\sim 1~cm^{-3}$ near the GC) WIM in the model, the scattering measure of
which falls down linearly by almost an order of magnitude from the GC to within
1$^{\circ}$. This ionized gas appears to be uniformly distributed in $l$ and
$b$.
A schematic diagram of the model is shown in Fig. 2.

\begin{figure}[h]
{\includegraphics[width=0.9\textwidth,clip=true,angle=270,totalheight=15cm]{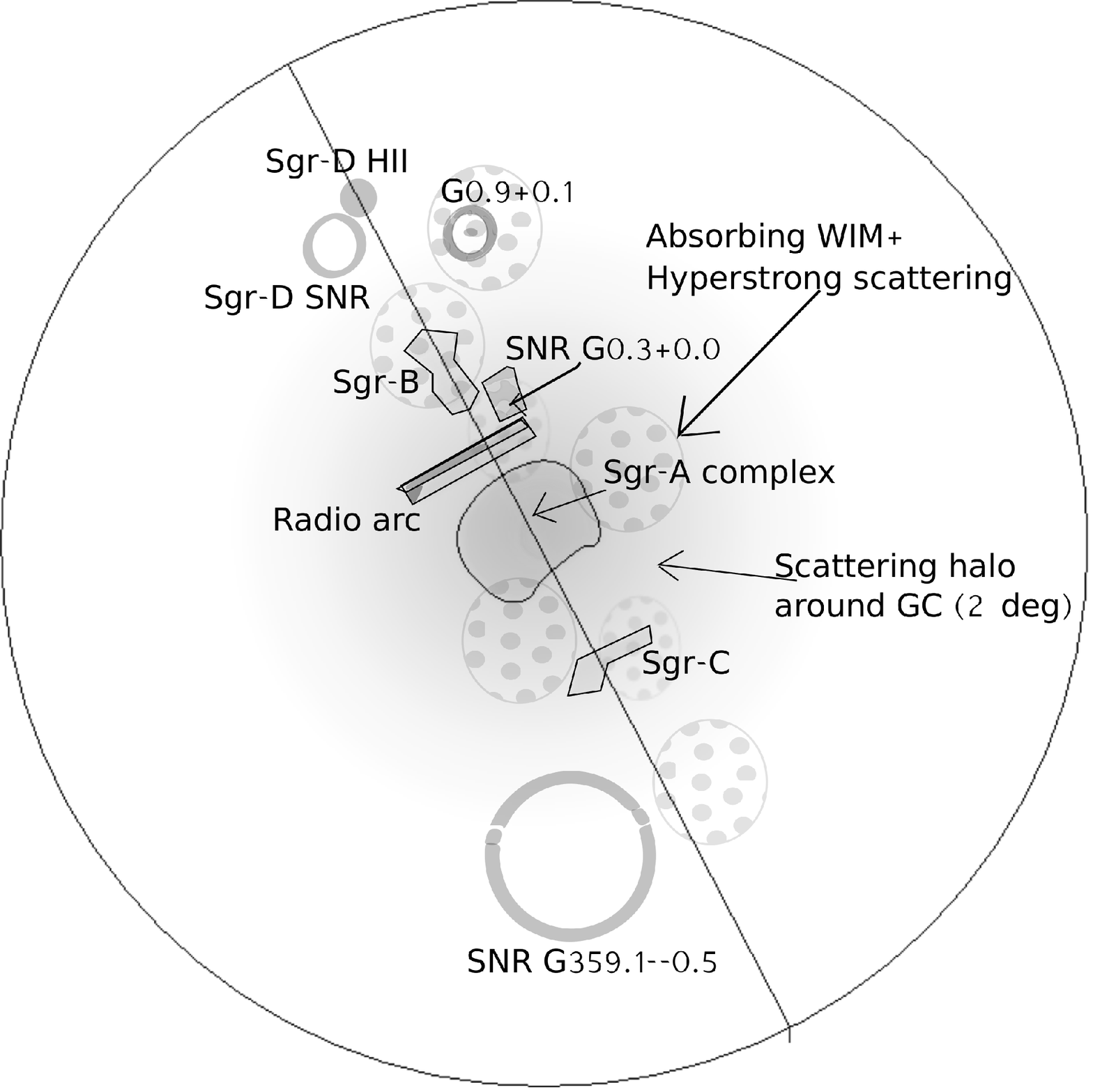}}
\caption{A schematic diagram of different components of the WIM near the GC
with the well known extended radio sources in the region. The central halo of
diameter of 2 degrees centered on the GC with a radial decrease in gray scale
shows the scattering screen as described in this paper. The circular regions
with smaller dark blobs show the high emission measure WIM that causes
absorption and heavy scattering of radio waves at low radio frequencies.}
\end{figure}

\section{Conclusions}

Deep observations of the GC region at 0.154 and 0.255 GHz bands with the GMRT
have allowed detection of 62 small diameter sources, most of which appear to be
extragalactic. Their intrinsic sizes and scattering diameters at $\lambda =1~$m
were derived from multifrequency observations. Derived scattering diameter is
30$'' \pm 2''.3$ at the GC and goes down with a slope of 27$''\pm$2$''.5$ per
degree within about 1$^{\circ}$ from the GC.  Considering the scattering size towards
the sources seen within 1$^{\circ}$ from the GC, we estimate an electron
density of $\sim 1~$cm$^{-3}$ in the region as responsible for the scattering.

Low frequency free-free absorption towards the extended sources
within 0.6$^{\circ}$ from the GC shows high absorption
at 0.154 GHz. Electron density of the ionized gas responsible for absorption is
likely to have $n_e \sim 10~cm^{-3}$. The probability of encountering this
absorbing screen is found to be $\sim$90\% within the region.
Following the Hyperstrong scattering model (NE2001), this ionized medium would
cause heavy scatter broadening of the EG sources and they would be resolved out
in 1.4 GHz interferometric observations.  However, we find 8 out of 10 sources
expected are still present in the 1.4 GHz catalog. Ionized surfaces of
molecular clouds are thought to provide the required high electron density gas that 
causes scattering and low radio frequency absorption.
Emission from these clouds as traced by CS $J=1-0$ transition shows that these
clouds are located close to the Galactic plane with an width of $\sim
0.1-0.2^{\circ}$.  Since most of the extended Galactic objects discussed here
are close to the Galactic plane, they are mostly seen in absorption.  EG
sources observed towards $359.5^{\circ} < l <0.5^{\circ}$ and $-0.5^{\circ} <b
<0.5^{\circ}$ of the Galaxy are seen through the so called Hyperstrong
scattering screen, but most of them are not seen through the dense molecular
clouds near the GC and hence are not resolved out due to heavy scattering
through the molecular cloud.

\section*{Acknowledgment}

I thank the staff of GMRT that allowed these observations to be made. GMRT is
run by National Centre for Radio Astrophysics of the Tata Institute of
fundamental research. A. Pramesh Rao was involved with this project, and made
the 0.154 GHz observations with the GMRT. I thank him for his contribution
including useful comments in improving the manuscript. I also thank the
anonymous referee for useful comments in improving the quality of the paper and
Aritra Basu for going through the manuscript. This research has made use of the
NED which is operated by the Jet Propulsion Laboratory, California Institute of
Technology, under contract with the National Aeronautics and Space
Administration.


\begin{thebibliography}{30}
\expandafter\ifx\csname natexlab\endcsname\relax\def\natexlab#1{#1}\fi

\bibitem[{{Anantharamaiah} {et~al.}(1991){Anantharamaiah}, {Pedlar}, {Ekers},
  \& {Goss}}]{ANANTHARAMAIAH1991}
{Anantharamaiah}, K.~R., {Pedlar}, A., {Ekers}, R.~D., \& {Goss}, W.~M. 1991,
  MNRAS, 249, 262

\bibitem[{{Becker} {et~al.}(2010){Becker}, {Helfand}, {White}, \&
  {Proctor}}]{BECKER2010}
{Becker}, R.~H., {Helfand}, D.~J., {White}, R.~L., \& {Proctor}, D.~D. 2010,
  \aj, 140, 157

\bibitem[{{Bower} {et~al.}(2001){Bower}, {Backer}, \& {Sramek}}]{BOWER2001}
{Bower}, G.~C., {Backer}, D.~C., \& {Sramek}, R.~A. 2001, ApJ, 558, 127

\bibitem[{{Brogan} {et~al.}(2003){Brogan}, {Nord}, {Kassim}, {Lazio}, \&
  {Anantharamaiah}}]{BROGAN2003}
{Brogan}, C.~L., {Nord}, M., {Kassim}, N., {Lazio}, J., \& {Anantharamaiah}, K.
  2003, Astronomische Nachrichten Supplement, 324, 17

\bibitem[{{Cohen} \& {Davies}(1976)}]{COHEN1976}
{Cohen}, R.~J. \& {Davies}, R.~D. 1976, MNRAS, 175, 1

\bibitem[{{Cordes}(2004)}]{CORDES2004}
{Cordes}, J.~M. 2004, in Astronomical Society of the Pacific Conference Series,
  Vol. 317, Milky Way Surveys: The Structure and Evolution of our Galaxy, ed.
  D.~{Clemens}, R.~{Shah}, \& T.~{Brainerd}, 211

\bibitem[{{Howe} {et~al.}(1991){Howe}, {Jaffe}, {Genzel}, \&
  {Stacey}}]{HOWE1991}
{Howe}, J.~E., {Jaffe}, D.~T., {Genzel}, R., \& {Stacey}, G.~J. 1991, ApJ, 373,
  158

\bibitem[{{Intema} {et~al.}(2011){Intema}, {van Weeren}, {R{\"o}ttgering}, \&
  {Lal}}]{INTEMA2011}
{Intema}, H.~T., {van Weeren}, R.~J., {R{\"o}ttgering}, H.~J.~A., \& {Lal},
  D.~V. 2011, \aap, 535, A38

\bibitem[{{Kassim} \& {Frail}(1996)}]{KASSIM1996}
{Kassim}, N.~E. \& {Frail}, D.~A. 1996, MNRAS, 283, L51

\bibitem[{{Lang} {et~al.}(1999){Lang}, {Anantharamaiah}, {Kassim}, \&
  {Lazio}}]{LANG1999a}
{Lang}, C.~C., {Anantharamaiah}, K.~R., {Kassim}, N.~E., \& {Lazio}, T.~J.~W.
  1999, ApJL, 521, L41

\bibitem[{{Lang} {et~al.}(2010){Lang}, {Goss}, {Cyganowski}, \&
  {Clubb}}]{LANG2010}
{Lang}, C.~C., {Goss}, W.~M., {Cyganowski}, C., \& {Clubb}, K.~I. 2010, \apjs,
  191, 275

\bibitem[{{LaRosa} {et~al.}(2000){LaRosa}, {Kassim}, {Lazio}, \&
  {Hyman}}]{LAROSA2000}
{LaRosa}, T.~N., {Kassim}, N.~E., {Lazio}, T.~J.~W., \& {Hyman}, S.~D. 2000,
  AJ, 119, 207

\bibitem[{{Lazio} {et~al.}(1999){Lazio}, {Anantharamaiah}, {Goss}, {Kassim}, \&
  {Cordes}}]{LAZIO1999}
{Lazio}, T.~J.~W., {Anantharamaiah}, K.~R., {Goss}, W.~M., {Kassim}, N.~E., \&
  {Cordes}, J.~M. 1999, ApJ, 515, 196

\bibitem[{{Lazio} \& {Cordes}(1998)}]{LAZIO1998}
{Lazio}, T.~J.~W. \& {Cordes}, J.~M. 1998, ApJ, 505, 715

\bibitem[{{Lazio} \& {Cordes}(2008)}]{LAZIO2008}
{Lazio}, T.~J.~W. \& {Cordes}, J.~M. 2008, \apjs, 174, 481

\bibitem[{{Maron} {et~al.}(2000){Maron}, {Kijak}, {Kramer}, \&
  {Wielebinski}}]{MARON2000}
{Maron}, O., {Kijak}, J., {Kramer}, M., \& {Wielebinski}, R. 2000, \aaps, 147,
  195

\bibitem[{{Marti} {et~al.}(1998){Marti}, {Mirabel}, {Chaty}, \&
  {Rodriguez}}]{MARTI1998}
{Marti}, J., {Mirabel}, I.~F., {Chaty}, S., \& {Rodriguez}, L.~F. 1998, \aap,
  330, 72

\bibitem[{{Mereghetti} {et~al.}(1998){Mereghetti}, {Sidoli}, \&
  {Israel}}]{MEREGHETTI1998}
{Mereghetti}, S., {Sidoli}, L., \& {Israel}, G.~L. 1998, A\&A, 331, L77

\bibitem[{{Nord} {et~al.}(2004){Nord}, {Lazio}, {Kassim}, {Hyman}, {LaRosa},
  {Brogan}, \& {Duric}}]{NORD2004}
{Nord}, M.~E., {Lazio}, T.~J.~W., {Kassim}, N.~E., {et~al.} 2004, AJ, 128, 1646

\bibitem[{{Rodr{\'{\i}}guez-Fern{\'a}ndez} \&
  {Mart{\'{\i}}n-Pintado}(2005)}]{RODRIGUEZ2005}
{Rodr{\'{\i}}guez-Fern{\'a}ndez}, N.~J. \& {Mart{\'{\i}}n-Pintado}, J. 2005,
  A\&A, 429, 923

\bibitem[{{Roy}(2003)}]{ROY2003}
{Roy}, S. 2003, A\&A, 403, 917

\bibitem[{{Roy} \& {Pramesh Rao}(2004)}]{ROY2004}
{Roy}, S. \& {Pramesh Rao}, A. 2004, \mnras, 349, L25

\bibitem[{{Roy} \& {Pramesh Rao}(2006)}]{ROY2006}
{Roy}, S. \& {Pramesh Rao}, A. 2006, Journal of Physics Conference Series, 54,
  156

\bibitem[{{Roy} \& {Rao}(2009)}]{ROY2009}
{Roy}, S. \& {Rao}, A.~P. 2009, in Astronomical Society of the Pacific
  Conference Series, Vol. 407, Astronomical Society of the Pacific Conference
  Series, ed. {D.~J.~Saikia, D.~A.~Green, Y.~Gupta, \& T.~Venturi}, 267--+

\bibitem[{{Roy} {et~al.}(2005){Roy}, {Rao}, \& {Subrahmanyan}}]{ROY2005}
{Roy}, S., {Rao}, A.~P., \& {Subrahmanyan}, R. 2005, \mnras, 360, 1305

\bibitem[{{Taylor} \& {Cordes}(1993)}]{TAYLOR1993}
{Taylor}, J.~H. \& {Cordes}, J.~M. 1993, ApJ, 411, 674

\bibitem[{{Tsuboi} {et~al.}(1999){Tsuboi}, {Handa}, \& {Ukita}}]{TSUBOI1999}
{Tsuboi}, M., {Handa}, T., \& {Ukita}, N. 1999, ApJS, 120, 1

\bibitem[{{van Langevelde} {et~al.}(1992){van Langevelde}, {Frail}, {Cordes},
  \& {Diamond}}]{VANLANGEVELDE1992}
{van Langevelde}, H.~J., {Frail}, D.~A., {Cordes}, J.~M., \& {Diamond}, P.~J.
  1992, ApJ, 396, 686

\bibitem[{{Wang} {et~al.}(2006){Wang}, {Dong}, \& {Lang}}]{WANG2006}
{Wang}, Q.~D., {Dong}, H., \& {Lang}, C. 2006, \mnras, 371, 38

\bibitem[{{White} {et~al.}(2005){White}, {Becker}, \& {Helfand}}]{WHITE2005}
{White}, R.~L., {Becker}, R.~H., \& {Helfand}, D.~J. 2005, \aj, 130, 586

\end{thebibliography}

\end{document}